\documentclass[a4paper,superscriptaddress,twocolumn,prl]{revtex4}
\usepackage[pdftex]{graphicx}
\usepackage[]{natbib}
\usepackage{todonotes}

\begin{document}

\title{$80\hbar k$ Momentum Separation with Bloch Oscillations in an Optically Guided Atom Interferometer}

\author{G. D. McDonald}
\email{gordon.mcdonald@anu.edu.au}
\homepage{http://atomlaser.anu.edu.au/}
\author{C. C. N. Kuhn}
\author{S. Bennetts}
\author{J. E. Debs}
\author{K. S. Hardman}
\author{M. Johnsson}
\author{J. D. Close}
\author{N. P. Robins}

\affiliation{Quantum Sensors Lab, Department of Quantum Science, Australian National University, Canberra, 0200, Australia}

\date{\today} 

\begin{abstract}
We demonstrate phase sensitivity in a horizontally guided, acceleration-sensitive atom interferometer with a momentum separation of $80\hbar k$ between its arms. A fringe visibility of 7\% is observed. Our coherent pulse sequence accelerates the cold cloud in an optical waveguide, an inherently scalable route to large momentum separation and high sensitivity. We maintain coherence at high momentum separation due to both the transverse confinement provided by the guide, and our use of optical delta-kick cooling on our cold-atom cloud. We also construct a horizontal interferometric gradiometer to measure the longitudinal curvature of our optical waveguide.
\end{abstract}

\maketitle

%Large Momentum transfer and why Bloch usually fails.

\begin{figure}
\centering{}
 \includegraphics[width=1\columnwidth]{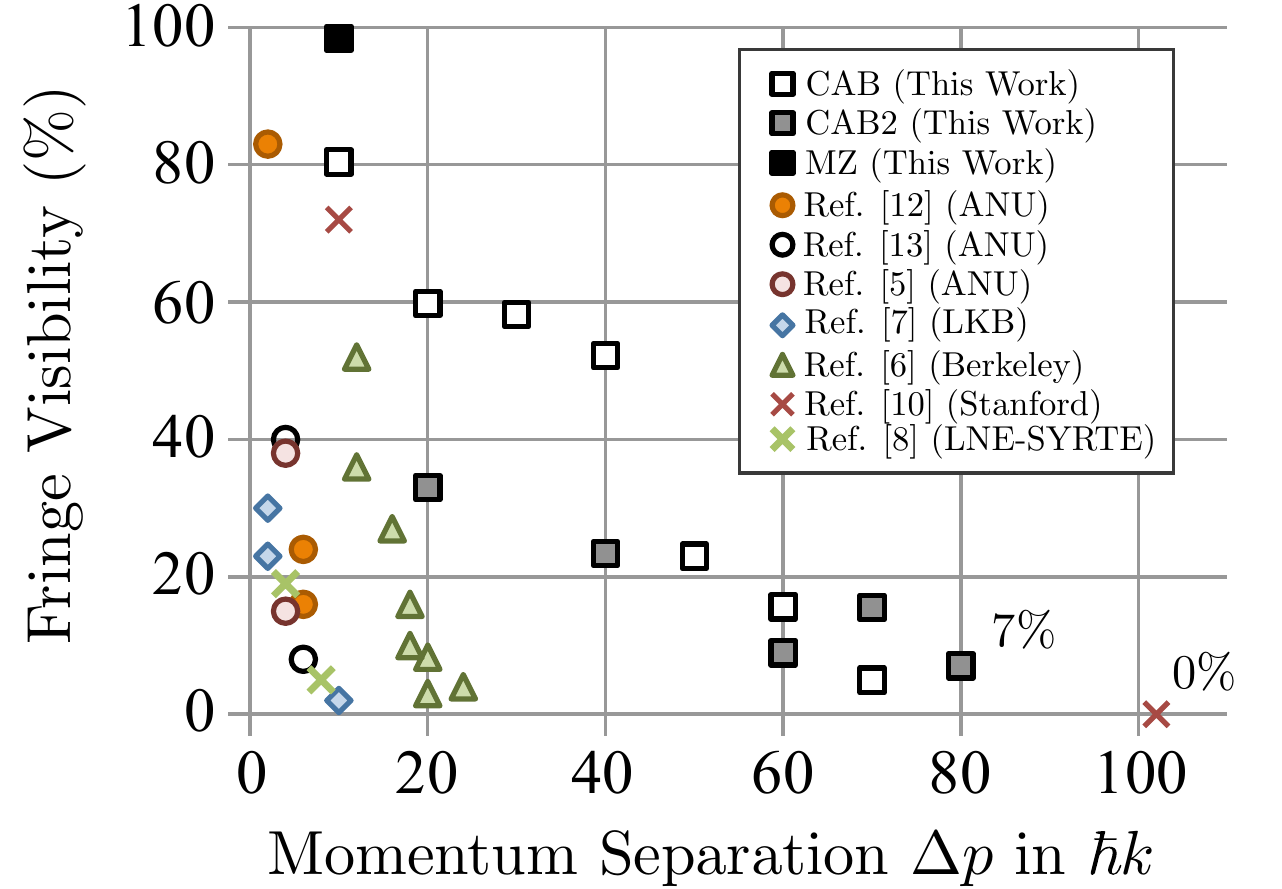}
 \caption{(Color online) Fringe visibility for various LMT accelerometer experiments \protect{\cite{SelfWaveguide,Muller24hk,CladeLMT,DoubleDiffraction,HMullerBlochBraggBloch,Kasevich102hk,TanhKasevich,DebsBECGrav,OurGravimeter}} as measured by the peak-to-peak amplitude of a sinusoidal fit to each fringe set. A standard $10\hbar k$ Mach-Zehnder and both the CAB and CAB2 pulse sequences used in this work (see text) are displayed for comparison. It should be noted that the fringe visibility of the interferometer with $\Delta p =102\hbar k$ in Ref.~\cite{Kasevich102hk} is zero, as phase noise prevented any phase measurement from being performed.}
 \label{LMTComparisonFig}
 \end{figure}

Cold-atom interferometers measure parameters of interest (for example an acceleration) by comparing the phase accumulated by an atom as it traverses either of two trajectories, known as the arms of the interferometer. Applications for such high-precision measurement devices include inertial sensing \cite{RobinsPhysicsReports}, gravitational wave detection \cite{BramGrav}, measurements of the fine structure constant \cite{Alpha2011} and tests of general relativity \cite{MullerGravityTest}. The sensitivity of an atom-interferometric accelerometer is proportional to its enclosed space-time area. Therefore, a key technology to enable the next generation of these  devices is Large Momentum Transfer (LMT), in which the enclosed space-time area is enlarged by increasing the momentum difference $\Delta p$ of the two interferometer arms. Various configurations for an LMT interferometer have been demonstrated \cite{SelfWaveguide,DebsBECGrav,OurGravimeter,Muller24hk,HMullerBlochBraggBloch,CladeLMT,DoubleDiffraction,Kasevich102hk,TanhKasevich} (see Fig.~\ref{LMTComparisonFig}) with $\Delta p$ up to $102\hbar k$ \cite{Kasevich102hk}, $k$ being the wavevector of the light used to effect the transition. However, a direct measurement of the interferometric phase and hence the ability to make an acceleration measurement has proved elusive beyond $\Delta p=24\hbar k$ \cite{Muller24hk}.

Here we measure the interferometric phase in a Bloch oscillation-based optically guided LMT atom interferometer with a momentum separation of $\Delta p$ up to $80\hbar k$. We use this phase measurement to calculate the tilt of the waveguide with respect to gravity. We maintain a fringe visibility of 7\% at $80\hbar k$ separation, as measured by a sinusoidal fit to the data \footnote{Some authors refer to the spread of the data as the contrast of an interferometer, whether or not an interferometric phase can be discerned. For example in Ref.~\cite{Kasevich102hk} a contrast of 18\% is reported, but with so much phase noise as to prevent identification of a visible interferometric fringe (i.e. 0\% visibility), or make any measurement of interferometric phase. By this metric, our $80\hbar k$ interferometer has a contrast of 19\%.} which we attribute to both our narrow longitudinal velocity width after optical delta-kick cooling and the transverse confinement of the optical guide. We characterize the longitudinal curvature of our optical waveguide by constructing a gradiometer in the guide. We also demonstrate a single beamsplitter with $\Delta p = 510\hbar k$, and an exponential decay time for the atoms held in the optical waveguide of 3.3s, demonstrating the scalability of this approach to LMT.
 %Comparison of all acceleration sensitive LMT.
 
  \begin{figure}
\centering{}
  \includegraphics[width=1\columnwidth]{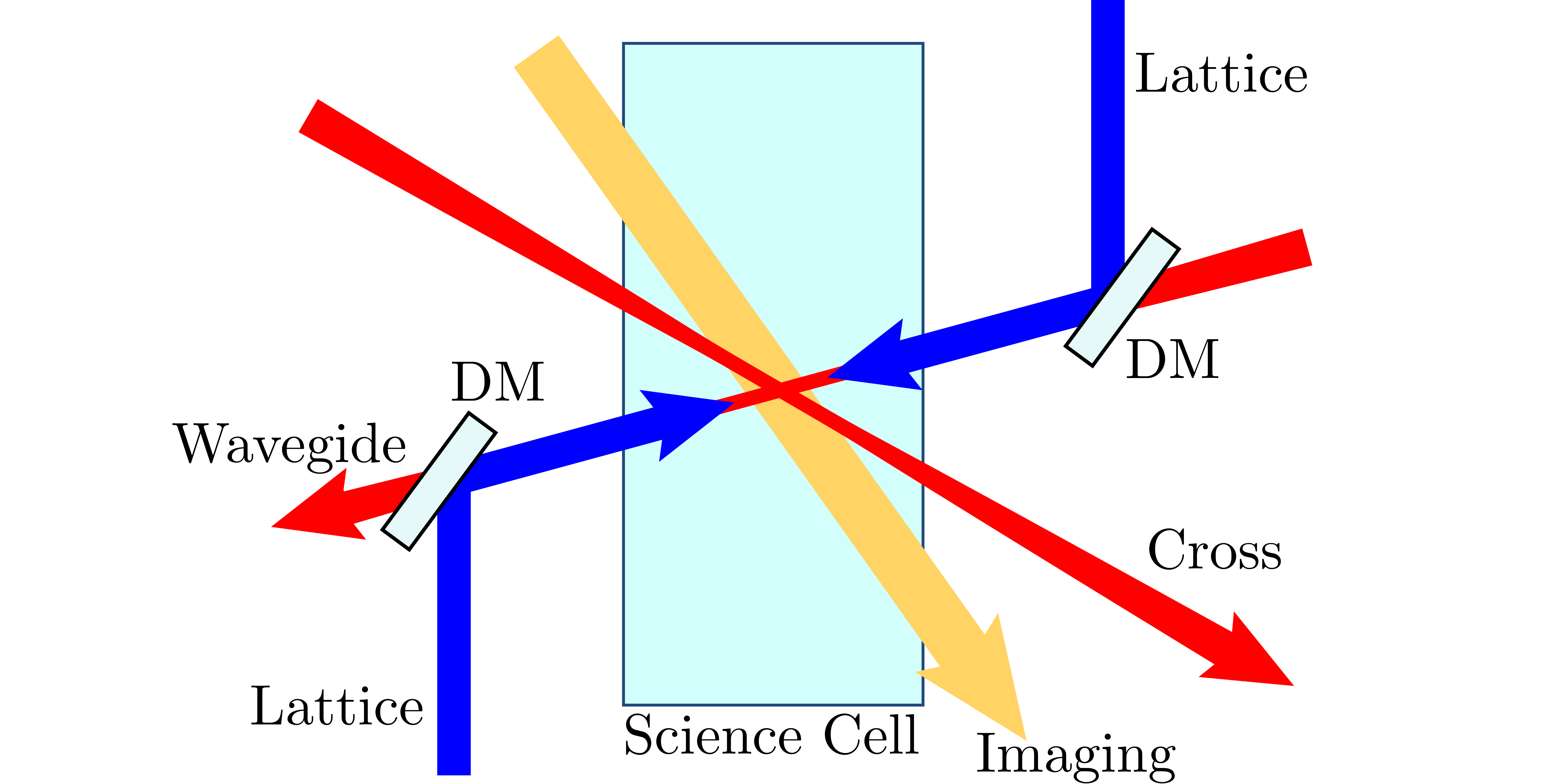}
 \caption{(Color online) We form a BEC in a cross beam dipole trap, before releasing it into the waveguide. After it has expanded for 125ms, we flash on the cross beam again to apply delta-kick cooling to the atoms. Our interferometer is formed by Bragg and Bloch pulses from counter-propagating beams aligned collinear with the waveguide beam.}
 \label{PotentialFig}
 \end{figure}
 %Focussing Calibration data and Bloch phase scan

\begin{figure}
\centering{}
 \includegraphics[width=1\columnwidth]{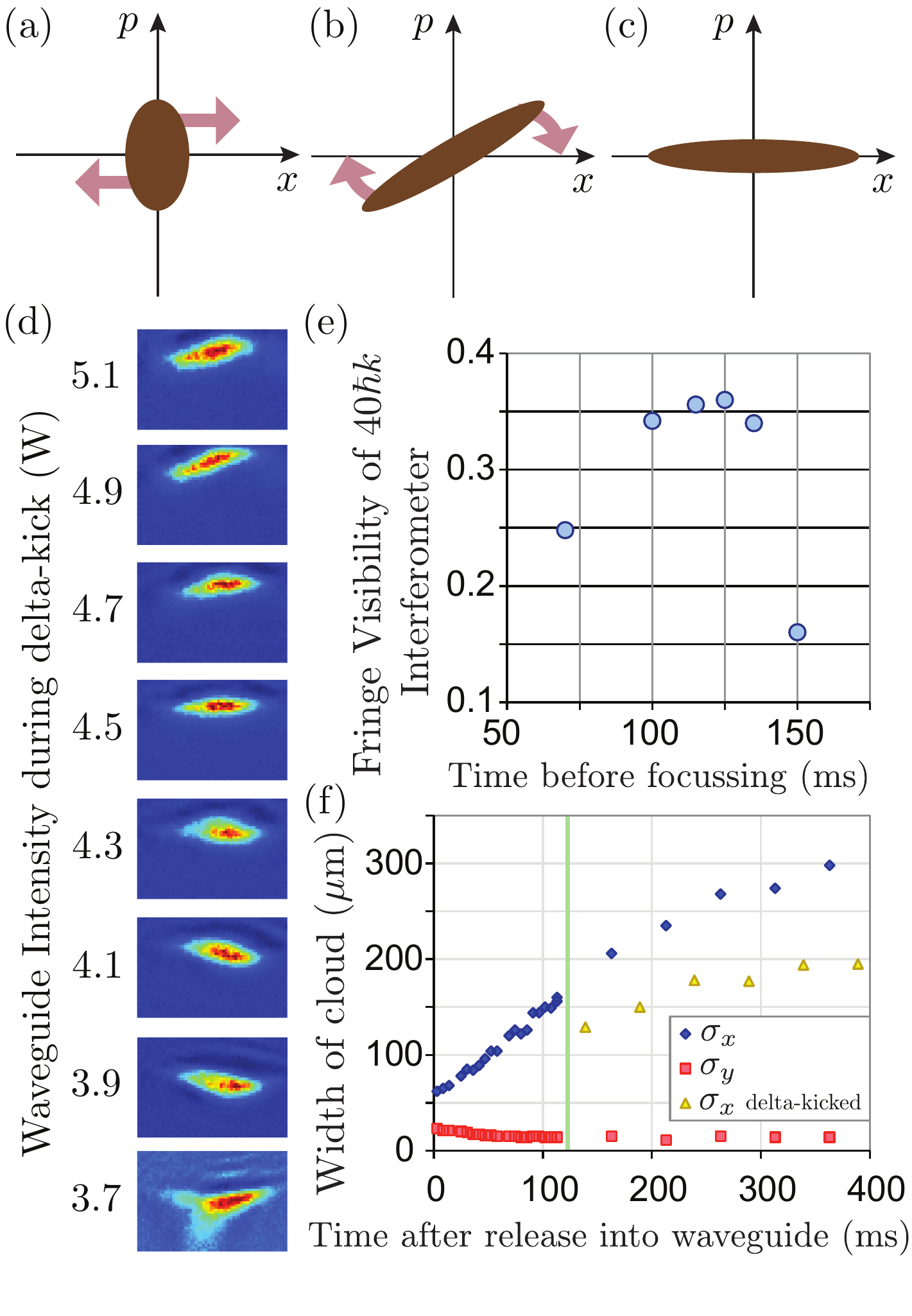}
 \caption{(Color online) (a)~Immediately after the condensate is released into the waveguide, it has a minimal spread in spatial extent and in momentum width. (b)~Over time, the cloud expands ballistically until momentum is well correlated with position along the guide. (c) Application of a harmonic potential for a short time (the delta-kick) reduces the momentum spread of the cloud, which now has a larger spatial extent. (d)~It is important to adjust the waveguide intensity so that the delta-kick cross-beam pulse is applied symmetrically over the cold cloud, otherwise transverse oscillations will occur in the guide. Here we show this adjustment process, in which the cloud is photographed a certain time after delta-kick cooling. The cloud is observed to oscillate if the waveguide power is either side of $4.5$ W. At 3.7 W we can see atoms falling out of the guide as they oscillate. (e) We optimize our delta-kick cooling by looking at the fringe visibility (as measured by a sinusoidal fit) of a $40\hbar k$ interferometer. We find our best visibility when our cross-beam flashes on 125ms after the atoms are released into the waveguide. (f) Of course, this collimation is imperfect; here we show the longitudinal width $\sigma_x$ of the cloud expanding after release as measured by the standard deviation of a gaussian fit, both with (yellow triangles) and without (blue diamonds) the cross beam flash at 125ms. Also shown is the transverse width $\sigma_y$ (  red squares). All widths are measured after an extra 22ms of ballistic expansion after the waveguide expansion time shown in (f).}
 \label{focussing}
 \end{figure} 
 		
% We make condensates, load into waveguide, focus

Our interferometric source is a $^{87}$Rb condensate formed by radio-frequency evaporation of atoms in their $\left|F=1, m_F=-1\right\rangle$ lower ground state in a hybrid magnetic/optical configuration \cite{HybridTrap} before transferring them into a crossed beam optical dipole trap (shown in Fig. \ref{PotentialFig}). The cross beam is sourced from a 2 nm line-width metal cutting laser operating at $1090$ nm, while the waveguide beam is a single frequency laser with 1 MHz line-width operating at $1064$ nm. The crossed dipole beams are adiabatically ramped down from 12 W each to 4.3 W and 175 mW respectively over 3 s which further evaporatively cools the atoms, producing a BEC of $2\times10^{6}$ atoms with a repetition rate of 2.5/min. We measure the axial trap frequency just before release into the waveguide to be {9 Hz}, by measuring the momentum oscillations after a $2\hbar k$ Bloch acceleration. Similarly, by misaligning the Bragg beams and giving a kick after release into the waveguide we measure the transverse (radial) frequency to be 60 Hz. As the cross beam is adiabatically ramped off, the waveguide intensity is increased back to 4.5 W so as to hold the atoms against gravity. We then wait a time $t_f$ for the atoms to expand in the guide, during which time they convert their mean-field energy into the kinetic energy of their velocity spread \cite{DebsBECGrav,SelfWaveguide} and then expand further until the position along the guide is well correlated with momentum (See Fig.~\ref{focussing} (a) and (b) ). Now the dipole cross beam is flashed on again for 2ms, providing an approximately harmonic potential which decelerates the faster atoms. This is an example of delta-kick cooling, which has been employed previously using Quadrupole-Ioffe magnetic traps \cite{BECDropTower,deltaKickCooling}, but has not yet been reported for an optically generated harmonic potential. This technique effectively rotates the ellipse describing position-momentum correlation along the waveguide so as to have minimal spread in momenta across the cloud (See Fig.~\ref{focussing} (c) ). In the case of a non-point source of atoms, and due to both the finite-size and anharmonicity of our dipole cross beam potential our delta-kick cooling is not ideal, so in practice we calibrate the process by measuring the fringe visibility of a $40\hbar k$ interferometer performed after various configurations. We find $t_f=125$ ms for our optimal delta-kick cooling configuration. Figure~\ref{focussing}~(e) shows this calibration and demonstrates that a narrow-momentum-width atom source is critical for reasonable fringe visibility in an LMT interferometer. By a fit to the expanding delta-kick-cooled cloud width (Figure~\ref{focussing}~(f), yellow triangles) we see that our interferometric atom source now has a momentum with of $0.05\hbar k$.
 
 %Pulse sequence diagrams

\begin{figure}
\centering{}
 \includegraphics[width=1\columnwidth]{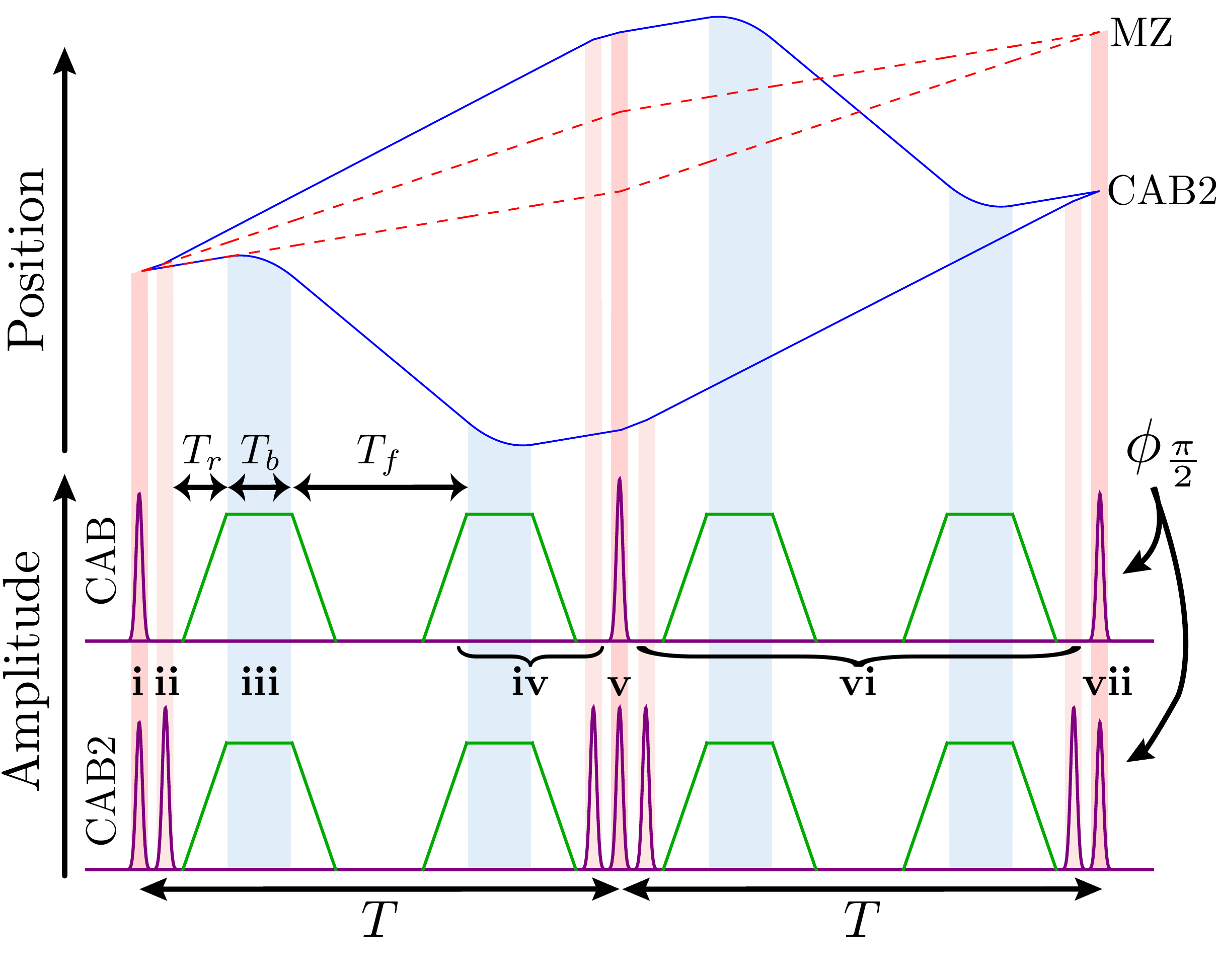}
 \caption{(Color online) Top: The interferometer paths over time for both our $10\hbar k$ MZ (red, dashed) and our $80\hbar k$ CAB2 (blue, solid) interferometers. Middle: This is our CAB interferometry sequence, in which a $10\hbar k$ Bragg (purple) MZ has its arms further momentum separated by up to $60\hbar k$ by Bloch oscillations (green). We present results for $2T=2$ms in this configuration. Bottom: This is our CAB2 sequence, which combines $20\hbar k$ sequential Bragg splitting with up to $60\hbar k$ Bloch lattice acceleration. For this sequence we present results for $2T=2.6$ms. The extra 0.6ms is to accommodate the four extra Bragg pulses included in the sequence.}
 \label{PulseSequenceFigure}
 \end{figure}

 %Bragg system
Our optical lattice laser setup has been described previously \cite{SelfWaveguide}. We have up to 50mW in each of two counter-propagating beams. These are aligned collinear with the waveguide in a two-step process. First, the small fraction of waveguide light which reflects off the dichroic mirror (see Fig.~\ref{PotentialFig}) is back coupled into the optical fibre which one of the lattice beams comes from. Secondly, the other lattice beam is coupled into the same optical fibre. The lattice beams are collimated with a full $\frac{1}{e^2}$ width of 1.85mm and detuned 105 GHz to the blue from the $\left|F=1\right> \rightarrow\left|F'=2\right>$ transition of the $D_2$ line in $^{87}\text{Rb}$, which keeps the number of spontaneous emissions below 1\% of our total atom number during our interferometric sequence. Arbitrary, independent control of the frequency detuning and amplitude of each beam is achieved using a direct digital synthesizer. Prior to our interferometer, a velocity selection Bragg pulse of $10\hbar k$ is used to isolate the portion of atoms ($\approx 80\%$) with a narrow momentum width $\sigma_p$ from those not properly cooled by our delta-kick process. For clarity, our Constant-Acceleration Bloch (CAB) interferometer sequences will be described in the frame of these velocity selected atoms, which are themselves moving at $10\hbar k$ with respect to the laboratory frame. Each part of the sequence is labeled with roman numerals corresponding to its depiction in Figure~\ref{PulseSequenceFigure}.

	 $\mathbf{i.}$ Our CAB (CAB2) sequence begins with a $2n\hbar k$ $\frac{\pi}{2}$ Bragg pulse with $n=5$ applied to the atoms to coherently split them into two momentum states, one in the initial $0\hbar k$ state, the other traveling at $10\hbar k$ ($\mathbf{ii.}$ followed by an extra $10\hbar k$ Bragg kick which is given to the faster $10\hbar k$ atoms, taking them to $20\hbar k$). $\mathbf{iii.}$ The $0\hbar k$ atoms are then loaded into a Bloch lattice of 10-20 recoil energies over a rise time of $T_r=110\mu$s which is accelerated in the other direction up to $2n_b \hbar k=-60\hbar k$ depending upon the final momentum separation desired, over a time $T_b=150\mu$s. $\mathbf{iv.}$ After a free evolution time of $T_f$, these accelerations are reversed, to bring the atoms in the lower arm back to the $0\hbar k$ (and the upper arm back to $10\hbar k$). $\mathbf{v.}$ A time $T$ after the initial $\frac{\pi}{2}$ pulse we apply a $\pi$ Bragg pulse to invert the two momentum states before $\mathbf{vi.}$ repeating the acceleration and deceleration sequence, which now acts upon the opposite arm of the interferometer. $\mathbf{vii.}$ After another period $T$, the two halves of the atomic wave packet are overlapped again and we apply a second $\frac{\pi}{2}$ pulse to interfere the two states. We allow these final states to separate, then switch off the waveguide to allow ballistic expansion for 8ms to avoid lensing of the imaging light by the narrow, optically dense cloud of atoms.  

Using absorption imaging we count the number of atoms in each spatially separated momentum state. To remove the effect of run-to-run fluctuations in total atom number, the relative atom number in the $0\hbar k$ state $N_{rel}=N_{0\hbar k}/(N_{0\hbar k}+N_{10\hbar k})$ is used. The final images are analyzed with a Fourier decomposition algorithm described previously \cite{SelfWaveguide}, to determine which parts of our final atomic density distribution are contributing to the interference. By scanning the laser phase $\phi_\frac{\pi}{2}$ of the final $\frac{\pi}{2}$ Bragg pulse, we obtain fringes in $N_{rel}$ which oscillate according to $N_{rel}=A\cos(\Phi_0+n\phi_{\frac{\pi}{2}})+c$ where $\Phi_0$, the phase shift which is sensitive to an external constant acceleration $a$, is given by \cite{T3PRLArxiv}

\begin{equation}
	\label{eq:phase}
	\Phi_0=2\left(n+n_{b}\cdot \frac{T_{b}+T_{f}}{T}\right)\cdot kaT^2\,.
\end{equation}

The Bloch lattice acceleration rate $|a_p|$ implicitly appears in Eq.~\ref{eq:phase}, because $|a_p|=2n_b \hbar k / m_{\text{Rb}}T_b$.
The maximum adiabatic acceleration rate $|a_p|$ increases quadratically with lattice depth~\cite{PeikBloch}, and therefore also increases quadratically with available laser power. This means that a Bloch-based configuration such as the CAB sequence can achieve a larger sensitivity for a given laser power than an equivalent sequential-Bragg configuration~\cite{Kasevich102hk}, in which the momentum transferrable in each Bragg diffraction pulse increases as the square root of the available laser power~\cite{SzigetiBragg}. In practice the lattice depth is limited because it must not bind the other ``non-resonant" arm of the interferometer~\cite{CladeLMT}. By using additional sequential $10{\hbar}k$ Bragg $\pi$ pulses, it is possible to increase each Bragg splitting to an effective $\Delta p=20{\hbar}k$ split for our CAB2 sequence, as opposed to a $\Delta p=10\hbar k$ Bragg split in our CAB sequence. In this way it is possible to avoid unintentionally binding the other arm of the interferometer while the lattice depth is increased, so as to achieve a higher Bloch lattice acceleration rate and a higher total momentum separation.

 %Actual Data
\begin{figure}
\centering{}
 \includegraphics[width=1\columnwidth]{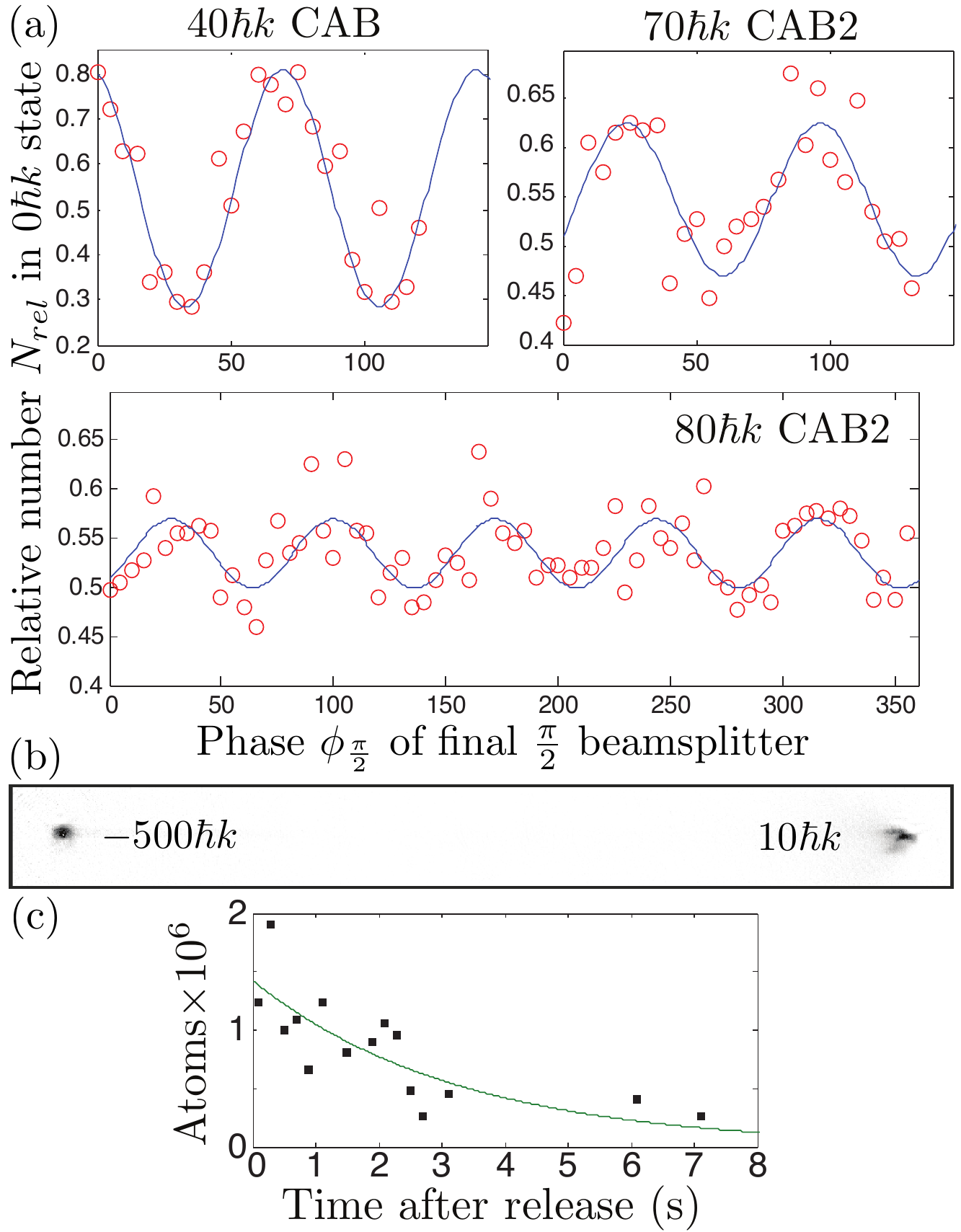}
 \caption{(Color online) (a)~Interferometric fringes recorded by scanning the laser phase $\phi_\frac{\pi}{2}$ of the final recombination $10\hbar k$ $\frac{\pi}{2}$ pulse. Measured data for $N_{rel}$ (red circles) and a sinusoidal fit (blue line) of the form $N_{rel}=A\cos(\Phi_0+n\phi_{\frac{\pi}{2}})+c$. The CAB sequence with $40\hbar k$ separation has $T=1$ms and fringe visibility $2A$ of 52\%, while the CAB2 sequence with $70\hbar k$ ($80\hbar k$) has $T=1.3$ms and a fringe visibility $2A$ of 16\% (7\%). Each data point in the $80\hbar k$ fringe is the average of three runs of the experiment. (b)~An absorption image of a BEC which has been split by $\Delta p=510\hbar k$ by the CAB beamsplitter. This was limited only by the size of the absorption image, as the clouds are separated by $\approx8$mm at the time of the image. (c)~Atoms remaining in the waveguide is plotted against hold time, showing an exponential decay time of 3.3s.}
 \label{fringes}
 \end{figure}
 
 In Fig.~\ref{LMTComparisonFig}, we show the fringe visibility we have observed for various interferometer configurations: a standard $\Delta p=10\hbar k$ MZ with 98\% visibility, our CAB sequence with the total momentum separation up to $\Delta p=70\hbar k$, and our CAB2 sequence with $\Delta p$ up to $80\hbar k$. We see that the CAB sequence, with its initial $10\hbar k$ Bragg splitting before the Bloch lattice is applied, decays to zero fringe visibility at a lower $\Delta p$ than our CAB2 sequence, which has an initial $20\hbar k$ sequential Bragg splitting. This result is in agreement with our earlier discussion about $|a_p|$ and lattice depth.
 
\begin{figure}
\centering{}
 \includegraphics[width=1\columnwidth]{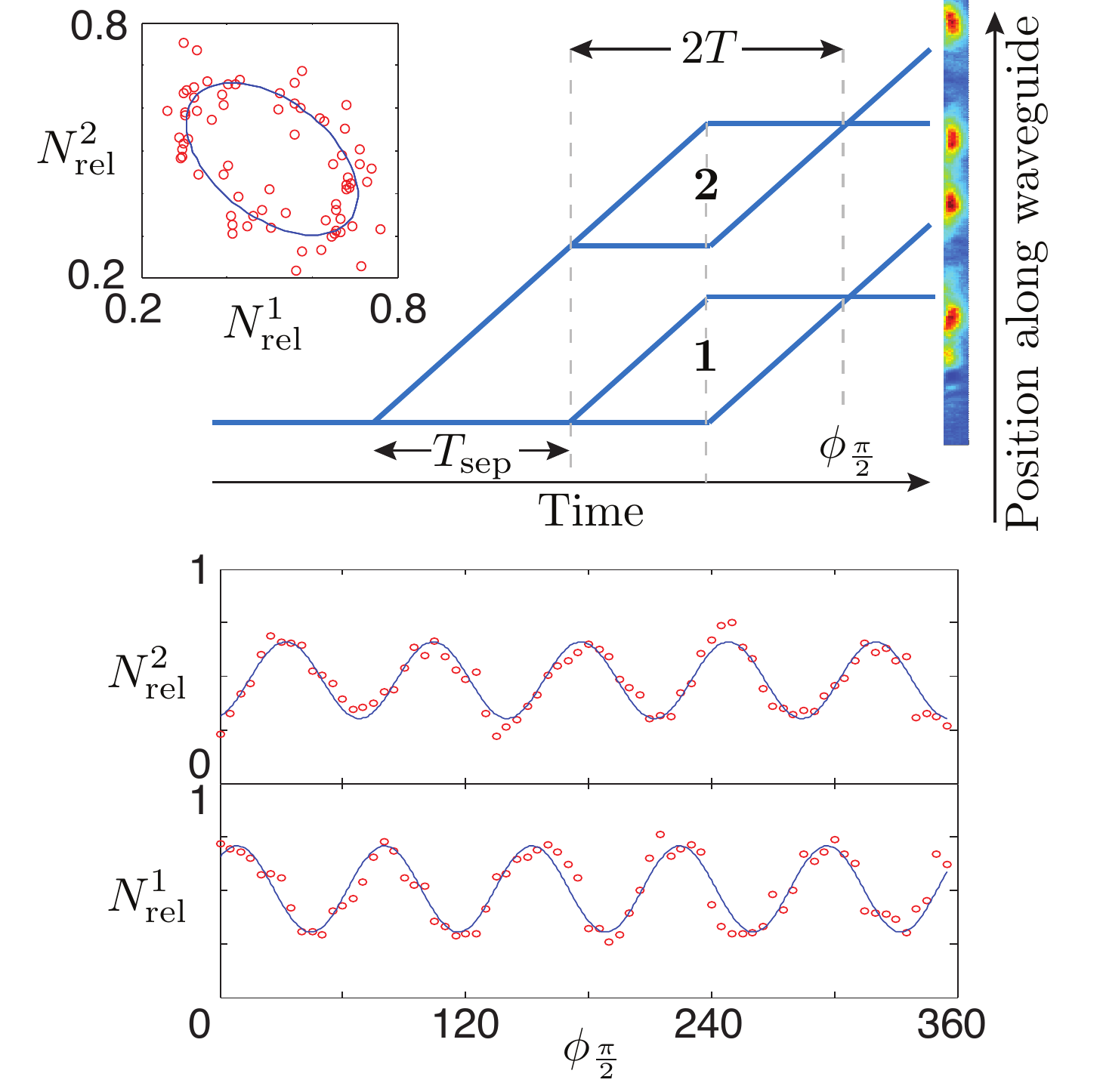}
 \caption{(Color online) Top: The atomic trajectories (not to scale) which make up the gradiometer. Four distinct states are seen which are derived from two spatially separate interferometers. Bottom: The relative atom number $N_{\text{rel}}^i$ in the $0\hbar k$ state from the $i$th interferometer is plotted while scanning the common phase $\phi_\frac{\pi}{2}$ of the last $\frac{\pi}{2}$ pulse. A fit to each sinusoid gives the acceleration-sensitive phase of each interferometer. Inset: When the two fringes are plotted parametrically against one another, an ellipse is generated from which the phase difference of the two interferometers can also be determined.}
 \label{gradiometer}
 \end{figure}
 
 %Results
 The maximum momentum separation we have achieved (while still being directly sensitive to phase) is $\Delta p=80\hbar k$, with a visibility of $7\%$ at $2T=2.6$ ms as seen in Fig.~\ref{fringes}~(a). The acceleration measured from this data is $a=3.1(1)\times10^{-3}$ m/s$^2$ from 146 runs of the experiment, which corresponds to that part of the waveguide being tilted $0.31(1)\,$mrad away from horizontal. Our best acceleration sensitivity of $7\times10^{-3} \,\text{m/s}^2\,\text{Hz}^{-1/2}$ is achieved at $\Delta p=70\hbar k$, also shown in Fig.~\ref{fringes}~(a). We have reduced phase noise as compared to measurements in the same laboratory \cite{SelfWaveguide,DebsBECGrav} by installing passive vibration isolation on both the science table and the laser table \footnote{We use closed air pods adjusted with a bicycle pump.}, along with the removal of all electronic equipment to an adjoining room to limit acoustic vibrations and electronic noise. 
 
  %Why can we get large Bloch accelerations? 
  
  We speculate that it is our use of an optical waveguide which allows us to achieve high Bloch accelerations without a drastic loss of coherence. There are three mechanisms we propose for this. First, since the atoms are transversely confined, they sample only a small segment of the comparatively much larger optical lattice beams so any spatial wavefront distortion due to an imperfect lattice beam mode is common to the whole interferometer. Secondly, the transverse confinement during Bloch acceleration allows for better mode matching at the final recombination pulse as compared with the use of a Bloch lattice in free space. Lastly, our use of an optical waveguide to support against gravity allows us to reduce the longitudinal velocity width via our optical delta-kick cooling. 
  
  Because the optical waveguide is formed at the shallow focus of a gaussian laser beam, it will have some curvature over the scale of the Rayleigh length. To measure the trapping frequency in the longitudinal direction (which is ideally zero), we construct a $\Delta p=10\hbar k=m_{\text{Rb}}v_{\text{rec}}$ gradiometer from two interferometers with a spatial extent $T v_{\text{rec}}=0.1$mm which are separated by a distance $x_2-x_1=T_{\text{sep}} v_{\text{rec}}=1$mm, as shown in Fig.~\ref{gradiometer}. Since the difference between the measured accelerations in each interferometer can be related to the trapping frequency $\omega$ of the waveguide curvature by $a_2-a_1=\omega^2(x_2-x_1)$, using Eq.~\ref{eq:phase} we can relate the phase difference $\Phi_2-\Phi_1$ between the two interferometers to the trapping frequency by
  
  \begin{eqnarray}
    \omega=\frac{1}{2nkT}\sqrt{\frac{m_{\text{Rb}}(\Phi_2-\Phi_1)}{\hbar T_{\text{sep}}}}\,.
  \end{eqnarray}
We calculate the longitudinal waveguide frequency by this method to be $\omega=2\pi\cdot 0.22(2)$~rad/s, using the data shown in Figure~\ref{gradiometer}. This compares well with a calculated estimate based upon the waveguide beam characteristics of $\omega=2\pi\cdot 0.18$~rad/s.

 %Look to the future

  One application of this system to high-sensitivity inertial sensing is in creating an accelerometer in which the acceleration sensitive phase scales as $T^3$, as opposed to a typical MZ which scales with $T^2$, or a Ramsey-Bord\'e configuration in which sensitivity scales with $T$. This is detailed in Ref.~\cite{T3PRLArxiv}. In an attempt to explore the boundaries of these kinds of LMT interferometer, we have constructed a $\Delta p=510\hbar k$ beamsplitter according to the CAB sequence, limited only by the size of the absorption image, and this is displayed in Fig.~\ref{fringes}~(b). In fact, Bloch lattices have been used to accelerate cold clouds by up to several thousand photon recoils~\cite{measureginalattice,Alpha2011,Alpha2008,CladeBounce,Ferrari8000hkBloch} but these configurations have no momentum separation between interferometric states, $\Delta p=0$. Future enhancement is also unaffected by hold time as we observe an exponential decay time of 3.3s for atoms held in our wavguide, as shown in Fig.~\ref{fringes}~(c).
 
 There are numerous avenues for future research in this system. By imaging a cold atom interferometer at the quantum-projection-noise limit \cite{DoeringQPNLimit} we can investigate large-atom-number squeezing directly via spatial overlap of the two states \cite{JohnssonSqueezingSelfInt,HaineNumberSqueezing,Esteve2008Squeezing}. The ability to hold all magnetic sub-states in the same waveguide spatial mode with an arbitrary, constant magnetic field allows us to completely remove the self-interaction in such a system by setting the scattering length to zero. Our apparatus is designed to also produce BEC of $^{85}$Rb and manipulate the s-wave scattering length via an easily accessible Feshbach resonance at 155 G \cite{Altin85Machine,AltinBosenova}. This could allow Heisenberg-limited delta-kick cooling of our atomic source, reaching even narrower momentum widths. The system offers the possibility of superimposing multidimensional lattices onto the propagating atoms to investigate universality in a 1D Bose gas \cite{Kuhn2012Criticality,Kuhn2012Universality}, or create the atom-optic equivalent of photonic crystals.  

%Conclusion
In summary we have shown a MZ interferometer based upon a CAB2 sequence with a momentum separation of up to $80\hbar k$. We have achieved an acceleration sensitivity of $7\times10^{-3} \,\text{m/s}^2\,/\sqrt{\text{Hz}}$ and a tilt sensitivity of $18$~mrad$/\sqrt{\text{Hz}}$. We attribute our ability to achieve large momentum separation using Bloch acceleration to our use of an optical waveguide. A single beamsplitter of $\Delta p=510\hbar k$ was constructed to demonstrate the scalability of this method. We also constructed a gradiometer which was used to measure the curvature of our optical waveguide with a sensitivity $\sigma_\omega=$0.1 rad/s. As an indication of the possible sensitivity this device is capable of, we can look at the quantum-projection-noise-limited sensitivity of a single run of an acceleration sensor with this architecture. Taking a momentum separation of $500\hbar k$, an interrogation time of $T=$ 50 ms (limited by a vacuum system of length 10 cm), and using $2\times10^6$ atoms, the shot noise limited sensitivity is $1.3\times10^{-10}$ m/s$^2$. This is the same sensitivity as could be achieved with a $2\hbar k$ interferometer with $T=580$ ms in the same vacuum system.

The authors would like to thank Hannah Keal and Paul Altin for their experimental assistance. We gratefully acknowledge the support of the Australian Research Council Discovery program. The author C.C.N. Kuhn would like to acknowledge financial support from CNPq (Conselho Nacional de Desenvolvimento Cientifico e Tecnologico). The author J.E. Debs would like to acknowledge financial support from the IC postdoctoral fellowship program.

\bibliographystyle{apsrev_v2}
\bibliography{GMcDonald80hk}
\end{document}